\documentclass[11pt]{article}
\usepackage{amsmath, amsthm, amssymb}
\usepackage{geometry}
\usepackage{graphicx}
\usepackage{tikz}
\usepackage{enumitem}
\usepackage{tabularx}
\usepackage{tikz}
\usepackage{amsmath}
\usepackage{amsfonts}
\usepackage{booktabs}
\usepackage{url} 
\usepackage{enumitem}
\usetikzlibrary{arrows.meta, positioning}
\usepackage{multirow}
\title{A Coincidence of Wants Mechanism for Swap Trade Execution in Decentralized Exchanges}

\author{
Abhimanyu Nag\thanks{Currently at University of Alberta, Department of Mathematical and Statistical Sciences} 
\and 
Madhur Prabhakar\thanks{Previously at Council for Scientific and Industrial Research, Ministry of Science and Technology, Government of India}
\and
Tanuj Behl\thanks{Previously at EQ8 Network}
}

\date{July 2025}

\newtheorem{definition}{Definition}
\newtheorem{observation}{Observation}

\begin{document}

\maketitle

\begin{abstract}
We propose a mathematically rigorous framework for identifying and completing Coincidence of Wants (CoW) cycles in decentralized exchange (DEX) aggregators. Unlike existing auction based systems such as CoWSwap, our approach introduces an asset matrix formulation that not only verifies feasibility using oracle prices and formal conservation laws but also completes partial CoW cycles of swap orders that are discovered using graph traversal and are settled using imbalance correction. We define bridging orders and show that the resulting execution is slippage free and capital preserving for LPs. Applied to real world Arbitrum swap data, our algorithm demonstrates efficient discovery of CoW cycles and supports the insertion of synthetic orders for atomic cycle closure. This work can be thought of as the detailing of a potential delta-neutral strategy by liquidity providing market makers: a structured CoW cycle execution.
\end{abstract}

\section{Introduction}

Modern decentralized exchanges (DEXs) \cite{malamud2017decentralized} rely on a blend of liquidity infrastructure and order routing mechanisms to settle asset swaps in real-time, permissionless environments. While traditional Automated Market Makers (AMMs) \cite{mohan2022automated} offer continuous price discovery and passive liquidity provision, recent developments in intent-based markets \cite{chitra2024analysis} and batch settlement architectures have shifted focus toward minimizing slippage, avoiding miner extractable value (MEV) \cite{milkau2023maximal}, and internalizing flow between users. One of the most theoretically powerful and practically underutilized concepts in this area is the \textit{Coincidence of Wants} (CoW)\footnote{While `Coincidence of Wants' has been used before in DeFi literature, especially in the context of CoWSwap (Section \ref{relwork}), we will describe a different and more rigorous definition of such a trade in this paper}.

In economic terms, a coincidence of wants \cite{lewis2014coincidence} arises when two or more parties can fulfill each other's demands through direct or indirect barter without involving a central intermediary. In decentralized finance (DeFi) \cite{zetzsche2020decentralized}, this translates to a closed loop of swap orders \cite{adams2024costs} where the output asset of each trader exactly matches the input asset of another. Such cycles permit atomic settlement using atomic transactions \cite{lynch1988theory} enabling efficient trade execution even in fragmented markets.

This paper develops a mathematically rigorous framework for identifying, validating, and completing CoW cycles in swap DEXs. We extend the notion of CoW beyond simple trader-to-trader matches by incorporating Liquidity Providers (LPs) as intermediating agents. Our formulation introduces: a graph-theoretic representation of swap orders and asset flows, a matrix formulation of formal feasibility constraints based on oracle pricing and an algorithm to discover partial or complete CoW cycles and construct bridging orders that close value-imbalanced loops.

Unlike heuristic-based auction solvers in CoWSwap \cite{shi2025competitive}, our methodology guarantees value-neutral execution. The resulting CoW cycles conserve both asset quantities and USD-equivalent value, making them suitable for vault-based DEX architectures and programmable liquidity protocols \cite{gogol2024sok}.

By simulating our algorithm over real-world Arbitrum \cite{bousfield2022arbitrum} swap data, we show that even sparse, unstructured order flow can yield non-trivial CoW opportunities. In doing so, we highlight how internal liquidity matching, when guided by algebraic structure rather than volume based routing \cite{lladrovci2023cow}, offers superior capital efficiency and system composability. The paper proceeds as follows : Section \ref{relwork} talks a bit about the related work (especially CoWSwap) and our motivation for this study, Section \ref{CoW} formalizes swap orders and defines Coincidence of Wants (CoW) cycles using directed graphs and asset flows, Section \ref{Algo} introduces a general CoW detection and completion algorithm with a motivating example and formal proof, Section \ref{sim} applies the algorithm to real swap data from Arbitrum, illustrating both complete and incomplete CoW cycles and simulating bridging logic where needed and Section \ref{conc} concludes with insights into capital efficiency, protocol design implications, and future extensions for CoW-based settlement primitives.

\section{Related Work} \label{relwork}

The idea of Coincidence of Wants (CoW) has recently gained practical relevance through protocols that exploit internal order matching without resorting to AMMs. Most prominently, \textbf{CoWSwap}\footnote{\url{https://cowswap.exchange}} by CowDAO is a decentralized exchange protocol that performs batch auctions to discover CoW opportunities off-chain and settle them on-chain via a solver competition. Orders submitted to CoWSwap are aggregated, and solvers propose atomic settlement sets which can include direct matches between traders (i.e., CoW cycles), as well as partial fills through external liquidity. More protocol details and analysis can be found in \cite{lladrovci2023cow}.

CoWSwap differs from AMMs like Uniswap \cite{adams2021uniswap} or Balancer \cite{ottina2023balancer}, where liquidity is passively pooled and pricing is determined by a bonding curve. In contrast, CoWSwap performs \emph{order flow internalization}, optimizing trade execution by resolving CoW matches before accessing liquidity pools, reducing slippage and avoiding unnecessary asset routing. Recent extensions in Balancer protocol also explores similar mechanisms by batching trades and minimizing MEV.  CoWSwap leverages batch auctions \cite{zhang2025maximal} and uniform clearing prices to protect users from MEV attacks such as sandwiching. When possible, CoWSwap identifies CoW pairs or cycles among submitted orders and settles them directly without interacting with on-chain AMMs.

However, none of these protocols provide a fully formal, graph based framework for CoW feasibility, dollar-valued neutrality, or bridging order generation as introduced in this paper. Our contribution departs from existing implementations in the following ways:
\begin{itemize}
    \item We construct a formal graph-theoretic and linear-algebraic model of CoW cycles, including asset and dollar-valued transfer matrices which allows us to find CoW opportunities, even in incomplete order networks, and propose completion orders algebraically.
    \item We define exact feasibility constraints using trader-specific exchange rate bounds for partial fills and oracle pricing to prevent slippage and handle CoW cycles of arbitrary length.
    \item We introduce a general algorithmic pipeline for discovering CoW cycles, detecting imbalances, and generating synthetic bridging swap orders to complete partial cycles.
    \item We provide a complexity analysis of the full matching algorithm and illustrate partial fill strategies that maintain internal value balance and guarantee closure under value conservation, thereby extending far beyond the heuristics of CoWSwap's implementation.
\end{itemize}

This work complements the practical systems built by CoWSwap and similar batch auction mechanisms, but our framework surpasses CoWSwap by replacing empirical cycle-matching with a provably correct and extensible matrix-theoretic formulation.

\section{Coincidence of Wants in Swap Systems} \label{CoW}

For the concept to work, we focus only on swap orders. Let $\chi$ denote the universe of tradable assets in a decentralized exchange. We define two relevant subsets: $\chi_A \subset \chi$ as the set of assets offered by traders, and $\chi_B \subset \chi$ as the set of assets requested by those traders in return. Those assets are typically sourced from liquidity providers (LPs) via smart contract vaults.

\subsection{Swap Orders}

\begin{definition}[Swap Order]
Let $\mathcal{O} = \{o_1, o_2, \ldots, o_n\}$ be a finite set of swap orders. Each order $o_i \in \mathcal{O}$ is defined as a tuple:
\[
o_i = (A_i, B_i, a_i, b_i)
\]
where:
\begin{itemize}
    \item $A_i \in \chi_A$ is the asset the trader offers,
    \item $B_i \in \chi_B$ is the asset the trader requests,
    \item $a_i \in \mathbb{R}_{>0}$ is the quantity of $A_i$ supplied,
    \item $b_i \in \mathbb{R}_{>0}$ is the quantity of $B_i$ desired in return.
\end{itemize}
The interpretation is that a trader is willing to trade $a_i$ units of asset $A_i$ in exchange for $b_i$ units of asset $B_i$.
\end{definition}

\subsection{Coincidence of Wants Cycles}

We now formalize the structure by which a subset of swap orders can mutually satisfy each other's asset demands in a closed loop, without the need for additional liquidity.

\begin{definition}[Coincidence of Wants (CoW) Cycle]
Let $\mathcal{C} = (o_{1}, o_{2}, \ldots, o_{k}) \subset \mathcal{O}$ be an ordered sequence of swap orders. The sequence $\mathcal{C}$ forms a \emph{Coincidence of Wants (CoW) cycle} if, for every $j \in \{1, \ldots, k\}$, the following condition holds:
\[
B_j = A_{(j \bmod k) + 1}
\]
That is, the asset desired in order $o_j$ is exactly the asset offered in order $o_{j+1}$, with indices taken modulo $k$ to ensure cyclic closure as a directed cycle.
\end{definition}

Consider the following three swap orders:
\begin{align*}
    o_1 &: (A = \text{ETH},\ B = \text{USDC},\ a_1 = 1,\ b_1 = 3000) \\
    o_2 &: (A = \text{USDC},\ B = \text{ARB},\ a_2 = 3000,\ b_2 = 1500) \\
    o_3 &: (A = \text{ARB},\ B = \text{ETH},\ a_3 = 1500,\ b_3 = 1)
\end{align*}
The sequence $(o_1, o_2, o_3)$ forms a CoW cycle since the assets requested and offered follow the closed loop:
\[
\text{ETH} \to \text{USDC} \to \text{ARB} \to \text{ETH}
\]

\subsection{Vault-Resolved CoW Cycles}

For liquidity sourcing, we propose trades are executed via vaults collateralised by Liquidity Providers (LPs), who commit asset inventory for matching purposes. A CoW cycle can be routed through LP vaults such that net asset movement is zero for each vault across the cycle.

\begin{definition}[LP Vault-Resolved CoW Cycle]
Let $\mathcal{L} = \{L_1, \ldots, L_m\}$ denote a finite set of LP vaults. A CoW cycle $\mathcal{C} = (o_1, \ldots, o_k)$ is said to be \emph{LP Vault-Resolved} if:
\begin{enumerate}
    \item Each swap order $o_j$ is intermediated by a pair of vaults $(L^{\text{in}}_j, L^{\text{out}}_j)$, where $L^{\text{out}}_j$ provides the asset $B_j$ and $L^{\text{in}}_j$ receives asset $A_j$.
    \item The net asset flow into and out of each vault in the cycle is zero: for every asset $T \in \chi$, the total amount of $T$ received by any vault equals the total amount of $T$ it supplies in the cycle.
\end{enumerate}
This ensures that each $L_j$ ends up with the same inventory they started with, preserving their capital over the atomic execution of the cycle.
\end{definition}

Continuing from the previous CoW cycle $(o_1, o_2, o_3)$, assume the following LP vault assignments:
\begin{itemize}
    \item $L_1$ is a $\text{USDC}$ Vault
    \item $L_2$ is an $\text{ETH}$ Vault
    \item $L_3$ is a $\text{ARB}$ Vault
\end{itemize}
Now according to our example,
\begin{itemize}
    \item $o_1$: $L_1$ provides $\text{USDC}$, $L_2$ receives $\text{ETH}$,
    \item $o_2$: $L_3$ provides $\text{ARB}$, $L_2$ receives $\text{USDC}$,
    \item $o_3$: $L_2$ provides $\text{ETH}$, $L_3$ receives $\text{ARB}$.
\end{itemize}
Assuming exchange rates are ideal and asset quantities match perfectly, each LP ends with their original asset and quantity. The cycle executes atomically with no slippage or residual imbalance.

\begin{figure}[h]
\centering
\begin{tikzpicture}[
    node distance=2.5cm and 4cm,
    every node/.style={font=\small},
    trader/.style={rectangle, draw=black, thick, minimum width=2.4cm, minimum height=1.1cm, align=center},
    vault/.style={rectangle, draw=black!60, thick, minimum width=2.4cm, minimum height=1.1cm, align=center},
    arrow/.style={->, thick, >=Stealth}
]

\node[trader] (T1) at (0, 4) {Trader 1};
\node[trader] (T2) at (0, 2) {Trader 2};
\node[trader] (T3) at (0, 0) {Trader 3};

\node[vault] (L1) at (5, 4) {$L_1$ \\ USDC Vault};
\node[vault] (L2) at (5, 2) {$L_2$ \\ ETH Vault};
\node[vault] (L3) at (5, 0) {$L_3$ \\ ARB Vault};

\draw[arrow] (L1) -- (T1) node[midway, above, sloped, pos=0.5] {USDC};
\draw[arrow] (L2) -- (T3) node[midway, above, sloped, pos=0.3] {ETH};
\draw[arrow] (L3) -- (T2) node[midway, above, sloped, pos=0.7] {ARB};

\draw[arrow] (T1) -- (L2) node[midway, below, sloped, pos=0.3] {ETH};
\draw[arrow] (T2) -- (L1) node[midway, below, sloped, pos=0.7] {USDC};
\draw[arrow] (T3) -- (L3) node[midway, below, sloped, pos=0.5] {ARB};

\end{tikzpicture}
\caption{A bipartite LP-resolved CoW cycle: each trader receives assets from one vault and sends different assets to another. Asset labels are clearly positioned to avoid overlap.}
\end{figure}
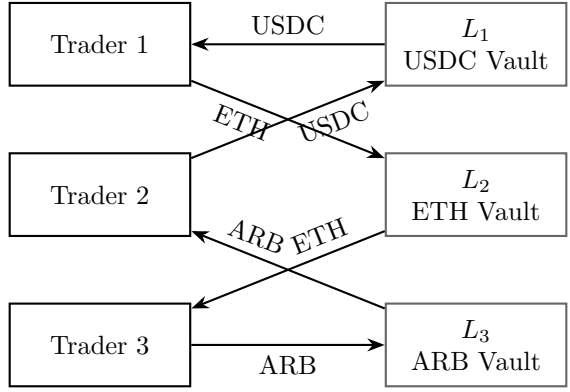

\subsection{Value Constraint Feasibility in CoW Cycles}

In practice, executed swaps deviate from ideal barter due to lack of matching quantities of swapped assets. To account for this, we introduce the notion of \textit{value feasibility} in CoW cycles.

We define \emph{acceptable exchange rate} $r_i > 0$, which specifies that the trader expects $r_i \cdot b_i$ units of asset $B_i$ in return.

Let $\mathcal{C} = (o_{i_1}, o_{i_2}, \ldots, o_{i_k})$ be a candidate CoW cycle of length $k$, with asset quantities $\{q_j\}_{j=1}^k$, where $q_j$ denotes the quantity of asset $A_{i_j}$ supplied in order $o_{i_j}$. The realized exchange rate for $o_{i_j}$ is defined as:
\[
\hat{r}_{i_j} := \frac{q_{j+1}}{q_j},
\]
where indices are taken cyclically, i.e., $q_{k+1} := q_1$.

The cycle $\mathcal{C}$ is said to be \textit{value-feasible} if all realized exchange rates meet or exceed the corresponding minimum requirements:
\[
\hat{r}_{i_j} \geq r_{i_j} \quad \text{for all } j \in \{1, \ldots, k\}.
\]

The aggregate surplus (or deficit) generated by the cycle is given by the product of the realized exchange rates:
\[
\prod_{j=1}^{k} \hat{r}_{i_j} = \frac{q_{2}}{q_1} \cdot \frac{q_3}{q_2} \cdots \frac{q_1}{q_k} = 1.
\]
Thus, ideally, the cycle is value-conserving. However, in practice, deviations due to quantity mismatch may occur, leading to:
\[
\prod_{j=1}^{k} \hat{r}_{i_j} \neq 1.
\]
In such cases, the cycle generates either a net surplus (arbitrage opportunity) or a deficit (loss), depending on whether the product exceeds or falls below unity. 

To enable partial fills and ensure minimal disruption in trade execution, we adopt a quantity selection strategy based on minimizing the \emph{minimum dollar value} transacted in the cycle:
\[
    \min_{j} \left( q_j \cdot p_{A_{i_j}} \right),
\]
where $p_{A_{i_j}}$ denotes the external reference price of asset $A_{i_j}$ (e.g. from an oracle). A price oracle ensures value neutrality by anchoring trades to external reference valuations, thereby preventing slippage-induced losses \cite{labadie2022impermanent} and preserving the integrity of LP vault balances. This mechanism also mitigates MEV risks by disallowing value-extractive settlements against outdated or imbalanced prices.

\section{Coincidence of Wants (CoW) Algorithm} \label{Algo}
This section formalizes the construction of an algorithmic framework for identifying and executing \textit{Coincidence of Wants} (CoW) cycles. The algorithm aims to both detect feasible CoW cycles and generate complementary swap orders that complete them. We begin with an motivating example.

\subsection{Intuitive Sketch of CoW Algorithm}

We consider a set of three swap orders that form a closed Coincidence of Wants (CoW) cycle. Ignoring market frictions and assuming all trades execute at quoted amounts (i.e., no slippage, zero protocol fee, and ideal matching), let the orders be specified as:

\begin{align*}
    o_1 &: (A = \text{ETH},\ B = \text{USDC},\ a_1 = 1,\ b_1 = 3000) \\
    o_2 &: (A = \text{USDC},\ B = \text{ARB},\ a_2 = 3000,\ b_2 = 1500) \\
    o_3 &: (A = \text{ARB},\ B = \text{ETH},\ a_3 = 1500,\ b_3 = 1)
\end{align*}
In this case, let $L_1$ be an $\text{ETH}$ Vault, $L_2$ is a $\text{USDC}$ Vault and $L_3$ is an $\text{ARB}$ Vault.
We represent this system as a directed graph $G = (V, E)$ where:
\begin{itemize}
    \item $V = \{\text{ETH}, \text{USDC}, \text{ARB}\}$ is the set of assets.
    \item $E = \{ (\text{ETH} \to \text{USDC}), (\text{USDC} \to \text{ARB}), (\text{ARB} \to \text{ETH}) \}$ are directed edges corresponding to swap intents.
\end{itemize}

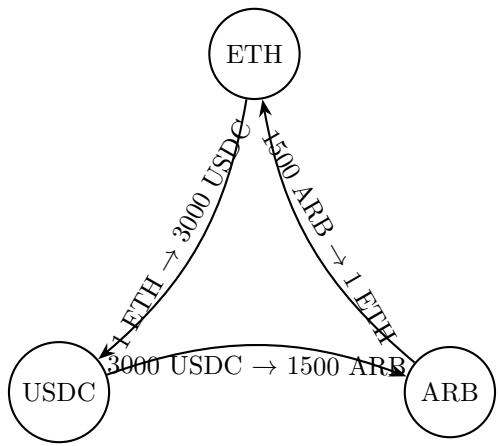
\begin{figure}[h]
\centering
\begin{tikzpicture}[->, >=Stealth, thick, node distance=3.5cm, every node/.style={font=\small}]
\tikzset{
    asset/.style={circle, draw=black, thick, minimum size=1.2cm, align=center}
}

\node[asset] (ETH) at (90:3) {ETH};
\node[asset] (USDC) at (210:3) {USDC};
\node[asset] (ARB) at (330:3) {ARB};

\draw[->] (ETH) to[bend left=20] node[midway, sloped, above] {1 ETH $\to$ 3000 USDC} (USDC);
\draw[->] (USDC) to[bend left=20] node[midway, sloped, below] {3000 USDC $\to$ 1500 ARB} (ARB);
\draw[->] (ARB) to[bend left=20] node[midway, sloped, above] {1500 ARB $\to$ 1 ETH} (ETH);
\end{tikzpicture}
\caption{Directed asset flow graph for a 3-order CoW cycle. Each edge denotes a desired asset exchange from source to target.}
\end{figure}

We encode the net asset flows of this CoW cycle using a matrix $M \in \mathbb{R}^{3 \times 3}$, where each row corresponds to a asset supplied by an order, and each column corresponds to the asset received. Let the asset ordering be vaults $\{L1,L2,L3\}$ as columns and positive entry implies addition of assets while a negative entry means a movement of assets out of the vault. Then:

\[
M = 
\begin{bmatrix}
\text{L1} & \text{L2} & \text{L3} \\
\hline
1 & -3000 & 0 \\
0 & 3000 & -1500 \\
-1 & 0 & 1500 \\
\end{bmatrix}
\]

We will call this \emph{Asset Transfer Matrix} or just \emph{Transfer Matrix}.

We now convert the previous CoW cycle into a dollar-valued representation using a price oracle such as chainlink or others. Let $p$ be a function that references oracle market price of assets in USD. Assume the following spot prices:
\begin{align*}
    p(\text{ETH}) &= \$3000 \\
    p(\text{USDC}) &= \$1 \\
    p(\text{ARB}) &= \$2
\end{align*}

Given the original orders:
\begin{align*}
    o_1 &: (1 \text{ ETH} \to 3000 \text{ USDC}) \\
    o_2 &: (3000 \text{ USDC} \to 1500 \text{ ARB}) \\
    o_3 &: (1500 \text{ ARB} \to 1 \text{ ETH}),
\end{align*}
we compute their dollar-equivalent trade values:
\begin{align*}
    o_1 &: (3000,\ -3000) \\
    o_2 &: (3000,\ -3000) \\
    o_3 &: (3000,\ -3000)
\end{align*}
Each tuple represents $(\text{value given}, \text{value received})$ in USD.

Let $V \in \mathbb{R}^{3 \times 3}$ be the dollar-valued asset transfer matrix, with the same ordering: $\{L1,L2,L3\}$. Then:
\[
V =
\begin{bmatrix}
3000 & -3000 & 0 \\
0 & +3000 & -3000 \\
-3000 & 0 & +3000 \\
\end{bmatrix}
\]

\subsubsection{Completing a CoW Cycle via New Order}
Consider a partial set of swap orders that do not yet form a complete Coincidence of Wants (CoW) cycle. As per this example we take:
\begin{align*}
    o_1 &: (A = \text{ETH},\ B = \text{USDC},\ a_1 = 1,\ b_1 = 3000) \\
    o_2 &: (A = \text{USDC},\ B = \text{ARB},\ a_2 = 3000,\ b_2 = 1500)
\end{align*}

The Asset transfer matrix $M \in \mathbb{R}^{2 \times 3}$ over assets $(\text{ETH}, \text{USDC}, \text{ARB})$ is given by:
\[
M =
\begin{bmatrix}
1 & -3000 & 0 \\
0 & 3000 & -1500 \\
\end{bmatrix}
\]

and the dollar-valued asset transfer matrix $V \in \mathbb{R}^{2 \times 3}$ is 
\[
V =
\begin{bmatrix}
3000 & -3000 & 0 \\
0 & 3000 & -3000 \\
\end{bmatrix}
\]
Let $N \in \mathbb{R}^{1 \times 3}$ denote the column-wise sum vector of $V$:
\[
N := \mathbf{1}^\top V = 
\begin{bmatrix}
3000 & 0 & -3000
\end{bmatrix}
\]

This vector denotes the net imbalance of the system: a deficit of $1$ ETH = \$$3000$, a surplus of $1500$ ARB = \$$3000$ and USDC is balanced.

Now to restore equilibrium and complete the CoW cycle, LPs or auxillary actors must introduce a new swap order $o_3$, called \emph{bridging order}, whose transfer vector is the negative of $N$. That is, we seek:
\[
V_{\text{new}} = -N = 
\begin{bmatrix}
+3000 & 0 & -3000
\end{bmatrix}
\]

This in denomination of original currency as matrix $M$ is :
\[
M_{\text{new}} = -N\cdot \text{diag}(p)^{-1} = 
\begin{bmatrix}
+1 & 0 & -1500
\end{bmatrix}
\]
where $$p = \begin{bmatrix}
+3000 & 0 & 0\\
0 & +1 & 0\\
0 & 0 & +2
\end{bmatrix}$$
This corresponds to a new swap order:
\[
o_3: (\text{ARB},\ \text{ETH},\ a_3 = 1500,\ b_3 = 1)
\]
The updated matrix becomes:
\[
M' = 
\begin{bmatrix}
-1 & +3000 & 0 \\
0 & -3000 & +1500 \\
+1 & 0 & -1500
\end{bmatrix}
\]

and 

\[
V' = 
\begin{bmatrix}
-3000 & +3000 & 0 \\
0 & -3000 & +3000 \\
+3000 & 0 & -3000
\end{bmatrix}
\]

Now, both the row-wise and column-wise sums of $V'$ are zero:
\[
\sum_{j=1}^{3} V'_{ij} = 0, \quad \sum_{i=1}^{3} V'_{ij} = 0
\]
for all $i,j$, indicating a closed and value-neutral cycle in dollar value terms. Meanwhile only the row sums of $M'$ are zero: 
\[
\sum_{i=1}^{3} M'_{ij} = 0
\]

\subsection{Formal Derivation of the Coincidence of Wants (CoW) Algorithm}

We now proceed to formalize the ideas introduced in the previous sections by constructing a mathematically rigorous algorithm for identifying Coincidence of Wants (CoW) cycles within a batch of swap transactions, as well as determining the minimal set of additional orders required to complete an incomplete CoW cycle. Throughout this section, we adopt the definitions and notational conventions previously established.
\subsubsection{Asset Transfer Matrix}
\begin{definition}[Asset Transfer Matrix]
    The resultant matrix $M \in \mathbb{R}^{k \times |\chi|}$ where each row $M_i$ corresponds to the net asset flow in vaults associated with order $o_i$:
\[
M_{i,j} := 
\begin{cases}
+a_i & \text{if } j = A_i \\
-b_i & \text{if } j = B_i \\
0 & \text{otherwise}
\end{cases}
\]
\end{definition}

Since each vault has a single asset, we will use vault and asset notation interchangeably. Let \( L = \{L_1, L_2, \ldots, L_n\} \) denote the set of all asset vaults appearing as columns in an asset transfer matrix \( M \in \mathbb{R}^{k \times |\chi| } \). Let \( p : \chi \to \mathbb{R}_{>0} \) be a price oracle assigning each asset \( L_j \in \chi \) a reference price \( p(L_j) \in \mathbb{R}_{>0} \).

Define the diagonal matrix \( P \in \mathbb{R}^{n \times n} \) as:
\[
P := \operatorname{diag}(p(L_1), p(L_2), \ldots, p(L_n))
\]
so that the \( j \)-th column of \( M \cdot P \) represents the dollar-denominated flow of asset \( L_j \), and each entry satisfies:
\[
(M \cdot P)_{i,j} = M_{i,j} \cdot p(L_j)
\]

\begin{definition}[Dollar-Normalized Transfer Matrix]
    Define the dollar-normalized transfer matrix $V \in \mathbb{R}^{k \times |\chi|}$ as:
\[
V_{i,j} = (M \cdot P)_{i,j} :=
\begin{cases}
+a_i \cdot p(A_i) & \text{if } j = A_i \\
-b_i \cdot p(B_i) & \text{if } j = B_i \\
0 & \text{otherwise}
\end{cases}
\]
\end{definition}

The total value row of $i$ is:
\[
\sum_{j=1}^{k} V_{i,j} = b_i \cdot p(B_i) - a_i \cdot p(A_i)
\]

\begin{definition}(Value-Feasibility)
    The cycle $\mathcal{C}$ is said to be \textit{value-feasible} if:
\[
\frac{p(B_i)}{p(A_i)} = \frac{a_i}{b_i} \quad \forall i \in \{1, \dots, k\}
\]
or equivalently, if each trader's expected return value is equal their contribution.
\end{definition}

If a trader’s order cannot be fully matched due to the absence of a complete CoW cycle or insufficient liquidity in the LP vaults, the protocol may execute a partial fill. This means only a fraction of the order is settled, but in a way that still fits into a valid and self-contained CoW cycle. 
Let $V \in \mathbb{R}^{k \times |\chi|}$ be the dollar-valued transfer matrix defined above. Then, under ideal execution and complete CoW closure, \( V \) satisfies the following structural invariants:

\begin{observation}[Row-wise value neutrality]
    For each agent $i \in \{1, \dots, k\}$, the net USD value exchanged is:
    \[
    \sum_{j=1}^{|\chi|} V_{i,j} = b_i \cdot p(B_i) - a_i \cdot p(A_i)
    \]
    Therefore, under value feasibility (i.e., $r_i = \frac{p(B_i)}{p(A_i)} = \frac{a_i}{b_i}$), we have:
    \[
    \sum_{j=1}^{|\chi|} V_{i,j} = 0
    \]
    implying that each agent receives in dollar terms exactly what they contribute.
\end{observation}

\begin{observation}[Column-wise asset value conservation]
    For each asset $x_j \in \chi$, let column $j$ represent the total dollar value flow of asset $x_j$ across all orders. Then:
    \[
    \sum_{i=1}^{k} V_{i,j} = 0 \quad \forall j \in \{1, \ldots, |\chi|\}
    \]
    This condition expresses global value neutrality across the system: the total dollar value of each asset supplied equals the total dollar value of that asset received. In other words, no dollar-denominated surplus or deficit is created in any individual asset.
\end{observation}

These two conditions jointly imply that the matrix $V$ lies in the null space of the all-ones row vector:
\[
\mathbf{1}^\top V = 0, \quad V \cdot \mathbf{1} = 0
\]
whenever all trades are executed at exact quoted rates and the CoW cycle is closed.
Let $N := \mathbf{1}^\top V \in \mathbb{R}^{|\chi|}$ be the column sum vector of $V$. Then:
\[
N_j = \sum_{i=1}^k V_{i,j} = \text{net dollar valued system-wide flow of  vault } j.
\]

\begin{observation}[Cycle Conservation]
     If $(o_1, \dots, o_k)$ form a closed CoW cycle with no leakages, then:
\[
N = 0 \in \mathbb{R}^{|\chi|}, \quad \text{i.e., } \sum_{i=1}^k V_{i,j} = 0 \ \forall j.
\]
\end{observation}

This expresses global asset conservation over the cycle: every asset supplied by one trader is exactly consumed by another.  Deviations from either of these constraints indicate the presence of surplus, slippage, or the need for bridging orders to restore balance.

\subsubsection{Cycle Feasibility and Bridging Orders}

Suppose a set of $k$ orders does not form a complete cycle. Let $M \in \mathbb{R}^{k \times |\chi|}$ be the associated transfer matrix, and compute the imbalance vector:

\begin{definition}[Imbalance Vector]
    \[
    N := \mathbf{1}^\top V \in \mathbb{R}^{|\chi|}
    \]
 If $N \neq 0$, then a complete CoW cycle is not formed. 
\end{definition}

 \begin{definition}[Bridging Order]
     Define a bridging order $o_{k+1}$ with net flow vector $V_{k+1} := -N$ so that:
\[
V' := 
\begin{bmatrix}
V \\
-N
\end{bmatrix}
\Rightarrow \mathbf{1}^\top V' = 0
\]
 \end{definition}
This enforces closure of the cycle in asset space.

\begin{observation}[Bridge Uniqueness]
    The vector $-N$ corresponds to a unique bridging order up to scaling that restores global asset neutrality. It is constructive: any incomplete path can be closed via an order $o^*$ whose asset flow is precisely $-N$.
\end{observation} 

\begin{observation}[Original Order Recovery]
    To recover the final bridging order in terms of the original asset transfer matrix $M$, we apply the following transformation : 
    $$M' = V' \cdot P^{-1} $$
\end{observation}

\subsection{Formal Algorithm (Mathematical Description)}

Let $\mathcal{O}$ be a set of swap orders. Then the algorithm proceeds as follows:

\begin{enumerate}
    \item Construct a directed asset graph $G = (V, E)$ with edge $(A_i \to B_i)$ for each $o_i \in \mathcal{O}$.
    \item For all simple cycles $\mathcal{C} \subseteq \mathcal{O}$ of length $\geq 2$:
    \begin{enumerate}
        \item Construct matrix $M_{\mathcal{C}}$ and compute $N = \mathbf{1}^\top M_{\mathcal{C}}$.
        \item If $N = 0$, test whether:
        \[
        b_i \cdot p(B_i) \geq a_i \cdot p(A_i) \quad \forall o_i \in \mathcal{C}
        \]
        \item If $N \neq 0$, define $o^*$ with flow $-N$ and test feasibility for $\mathcal{C} \cup \{o^*\}$.
    \end{enumerate}
    \item Return all feasible CoW cycles, complete or completed.
\end{enumerate}

\subsubsection{Computational Complexity of the CoW Algorithm}

Let $G = (V, E)$ be the directed asset flow graph induced by a set of swap orders $\mathcal{O}$, where each order $o_i = (A_i \to B_i)$ corresponds to a directed edge. Then:
\[
|E| = n, \quad |V| \leq 2n
\]

The first step is to enumerate all simple directed cycles $\mathcal{C} \subset G$ of length at most $k_{\max} \in \mathbb{N}$ (typically $k_{\max} \leq 4$). Let $c := |\{\mathcal{C} : \text{len}(\mathcal{C}) \leq k_{\max}\}|$.

Using Johnson’s algorithm \cite{johnson1977efficient}, all worst case cycles can be enumerated in time:
\[
\mathcal{O}\left((|V|^{2}\text{ log}|V| + |V||E|)(c + 1)\right)
\]
In the worst case, $c = \Theta(2^{|V|})$, but for sparse graphs (as in practice), $c \ll 2^{|V|}$.

For each candidate cycle \( \mathcal{C} \) of length \( k \), we perform the following computations. First, we construct the asset flow matrix \( M_{\mathcal{C}} \in \mathbb{R}^{k \times |\chi|} \) and compute the imbalance vector \( N = \mathbf{1}^\top M_{\mathcal{C}} \), both of which require \( \mathcal{O}(k) \) time. Next, we query the price oracle \( p : \chi \to \mathbb{R}_{>0} \) for the relevant assets involved in the cycle, requiring at most \( 2k \) lookups and therefore also taking \( \mathcal{O}(k) \) time. Finally, we evaluate the value-feasibility constraint \( b_i \cdot p(B_i) \geq a_i \cdot p(A_i) \) for each order \( o_i \in \mathcal{C} \), which again takes \( \mathcal{O}(k) \) time overall.

Across all $c$ candidate cycles:
\[
\text{Total time } = \mathcal{O}\left((|V|^{2}\text{ log}|V| + |V||E|)(c \cdot k_{\max} \right)
\]

The CoW algorithm has exponential worst-case complexity but may be tractable under realistic market conditions because the batch size of transactions will be small. Further work includes Preprocessing, heuristic pruning, and code optimization to ensure scalability in deployment.


\section{Simulation: Applying the CoW Algorithm to Real Swap Data} \label{sim}

To demonstrate the practical application of the Coincidence of Wants (CoW) mechanism, we conduct a simulation using real-world swap data through 1inch aggregator\footnote{\url{https://1inch.io/aggregation-protocol/}} collected from the Arbitrum blockchain on June 25, 2025. The dataset, shown in Table~\ref{tab:sample-swaps}, includes ten swap transactions routed through decentralized aggregators, capturing source and destination assets, along with the USD value of each trade. To evaluate the CoW framework in practice, we analyze a sample of swap transactions executed on the Arbitrum blockchain, routed through 1inch. We look for CoW cycles in batches of 10 transactions with an expiry time of 4 minutes (as batch sizes for the algorithm). The dataset contains 10 swap events, each recorded with the following attributes:
\begin{itemize}
    \item \texttt{time}: timestamp of the swap,
    \item \texttt{blockchain}: the chain of execution (all on Arbitrum),
    \item \texttt{tx\_hash}: transaction identifier,
    \item \texttt{amount\_usd}: USD value of the swap,
    \item \texttt{src\_asset\_symbol}, \texttt{dst\_asset\_symbol}: the input and output assets exchanged,
    \item \texttt{sender\_address}, \texttt{receiver}: the addresses initiating and receiving the swap.
\end{itemize}

\begin{table}[h]
\centering
\caption{Truncated Swap Transactions on Arbitrum (June 25, 2025)}
\label{tab:sample-swaps}
\begin{tabular}{lllll}
\toprule
\textbf{Time} & \textbf{Tx Hash (truncated)} & \textbf{USD Amount} & \textbf{Source asset} & \textbf{Destination asset} \\
\midrule
14:13 & 0x9b35...0cea & 14.01 & USDT & USDC \\
14:13 & 0xebfa...f8e7 & 2442.17 & ARB & USDC \\
14:13 & 0xf9d5...90b5 & 12.42 & WXM & USDC \\
14:12 & 0x0a78...d894 & 488.52 & USDC & ETH \\
14:12 & 0xb145...5f30 & 15.14 & USDT & DAI \\
14:11 & 0x7546...2291 & 219.61 & SolvBTC & USDC \\
14:11 & 0xe473...be5d & 0.34 & ETH & UNI \\
14:11 & 0x70da...e215 & 3.00 & USDT & ARB \\
14:10 & 0xa857...52ed & 0.24 & ETH & ARB \\
14:10 & 0x3350...360c & 20.24 & aArbWETH & ETH \\
\bottomrule
\end{tabular}
\end{table}

\subsection{Graph Construction and Cycle Detection}

We construct a directed graph $G = (V, E)$ from the dataset, where each vertex $v \in V$ represents a asset, and each directed edge $(A \rightarrow B) \in E$ corresponds to a swap order offering asset $A$ in exchange for asset $B$.

From Table~\ref{tab:sample-swaps}, the following edges are extracted:

\begin{center}
\small
\begin{tabular}{ll}
USDT $\rightarrow$ USDC, & ARB $\rightarrow$ USDC, \\
WXM $\rightarrow$ USDC, & USDC $\rightarrow$ ETH, \\
USDT $\rightarrow$ DAI, & SolvBTC $\rightarrow$ USDC, \\
ETH $\rightarrow$ UNI, & USDT $\rightarrow$ ARB, \\
ETH $\rightarrow$ ARB, & aArbWETH $\rightarrow$ ETH \\
\end{tabular}
\end{center}

Using Johnson’s algorithm, we identify all simple directed cycles. One notable 3-asset cycle is:

\[
\textbf{Cycle 1:} \quad \text{ETH} \rightarrow \text{ARB} \rightarrow \text{USDC} \rightarrow \text{ETH}
\]

This cycle is constructed from the following three transactions:

\begin{itemize}
\item ETH $\rightarrow$ ARB (Tx: 0xa857...52ed, \$0.24)
\item ARB $\rightarrow$ USDC (Tx: 0xebfa...f8e7, \$2442.17)
\item USDC $\rightarrow$ ETH (Tx: 0x0a78...d894, \$488.52)
\end{itemize}
We assume the oracle price as follows where $1 \text{ ETH }= 1500 \text{ ARB } = 1 \text{ aArbWETH } = 375 \text{ UNI } =3000 \text{ USDC}$

\subsection{Dollar-Normalized Matrix}

Since the swap data is already in USD, we build the dollar-normalized asset transfer matrix $V \in \mathbb{R}^{3 \times 3}$, where rows correspond to swap orders and columns to assets in the cycle (ETH, ARB, USDC). Given the lowest dollar value is \$0.24, this would correspond to a partial order fill across all orders in the cycle. Therefore the matrix $V$ would look like the following:

\[
M = \begin{bmatrix}
+0.24 & -0.24 & 0 \\
0 & +0.24 & -0.24 \\
-0.24 & 0 & +0.24 \\
\end{bmatrix}
\]
This implies that the corresponding CoW cycle is :

\[
\begin{aligned}
    o_1 &= (\text{ETH},\ \text{ARB},\ 0.24,\ -0.24) \\
    o_2 &= (\text{ARB},\ \text{USDC},\ 0.24,\ -0.24) \\
    o_3 &= (\text{USDC},\ \text{ETH},\ 0.24,\ -0.24)
\end{aligned}
\]

In original currency, the exact swaps are as follows :
\[
\fbox{
\parbox{0.85\linewidth}{
\[
\begin{aligned}
    o_1 &= (\text{ETH},\ \text{ARB},\ 0.00008,\ -0.12) \\
    o_2 &= (\text{ARB},\ \text{USDC},\ 0.12,\ -0.24) \\
    o_3 &= (\text{USDC},\ \text{ETH},\ 0.24,\ -0.00008)
\end{aligned}
\]
}
}
\]

\subsection{Completing a CoW Cycle from Arbitrary Orders}

We now apply the CoW algorithm to a different set of real-world swap orders drawn from Table~\ref{tab:sample-swaps}. These orders do not initially form a closed cycle, but we will compute the imbalance vector and demonstrate how to complete the cycle using bridging logic.

\subsubsection{Select Swaps from Dataset}

We identify the following two swap orders from Table~\ref{tab:sample-swaps}:

\[
\begin{aligned}
o_1 &= (\text{aArbWETH},\ \text{ETH},\ +20.24,\ -20.24) \\
o_2 &= (\text{ETH},\ \text{UNI},\ +0.34,\ -0.34) \\
\end{aligned}
\]

We can intuitively see a potential CoW cycle:

\[
\text{aArbWETH} \rightarrow \text{ETH} \rightarrow \text{UNI} \rightarrow \text{aArbWETH}
\]
\subsubsection{Construct Dollar-normalized asset Transfer Matrix}
Each swap is represented as a row in $V \in \mathbb{R}^{2 \times 3}$, with column order \texttt{[aArbWETH, ETH, UNI]}. Given the lowest dollar value is \$0.34, we can derive matrix $V$ as :

\[
V =
\begin{bmatrix}
+0.34 & -0.34 & 0 \\
0 & +0.34 & -0.34 \\
\end{bmatrix}
\]

which implies a partial fill for order $o_1$.
\subsubsection{Check asset Flow Balance}

We compute the imbalance vector \(N = \mathbf{1}^\top V\):

\[
N = 
\begin{bmatrix}
+0.34 & 0 & -0.34
\end{bmatrix}
\]

This reveals a small mismatch due to inconsistent USD valuations across swaps:
\begin{itemize}
    \item aArbWETH is short 0.34
    \item ETH is balanced
    \item UNI is in surplus of 0.34
\end{itemize}

\subsubsection{Adjust with a Bridging Order}

To complete the cycle and restore balance, we can introduce a single bridging order:

\[
o_4 = (\text{UNI},\ \text{aArbWETH},\ +0.34,\ -0.34)
\]

Now we extend the matrix:

\[
V' =
\begin{bmatrix}
0.34 & -0.34 & 0 \\
0 & +0.34 & -0.34 \\
-0.34 & 0 & +0.34
\end{bmatrix}
\]

Updated imbalance:

\[
\mathbf{1}^\top V' = [0, 0, 0]
\]

This confirms a value-neutral, closed CoW cycle among:

\[
\text{aArbWETH} \rightarrow \text{ETH} \rightarrow \text{UNI} \rightarrow \text{aArbWETH}
\]

with a small bridging order \((\text{UNI} \rightarrow \text{aArbWETH})\) required to reconcile minor pricing mismatch.

Therefore the CoW cycle is :

\[
\begin{aligned}
o_1 &= (\text{aArbWETH},\ \text{ETH},\ 0.34,\ -0.34) \\
o_2 &= (\text{ETH},\ \text{UNI},\ 0.34,\ -0.34) \\
o_3 &= (\text{UNI},\ \text{aArbWETH},\ 0.34,\ -0.34)
\end{aligned}
\]
and in the original currency this corresponds to :
\[
\fbox{
\parbox{0.85\linewidth}{
\[
\begin{aligned}
o_1 &= (\text{aArbWETH},\ \text{ETH},\ 0.0001133,\ -0.0001133) \\
o_2 &= (\text{ETH},\ \text{UNI},\ 0.0001133,\ -0.0425) \\
o_3 &= (\text{UNI},\ \text{aArbWETH},\ 0.425,\ -0.0001133)
\end{aligned}
\]
}
}
\]

\subsection{Simulated Output for CoW Cycle}
\subsubsection{Recognized CoW Cycle}
Here is the computer output of running the CoW algorithm on this sample dataset.
\begin{center}
\fbox{
\begin{minipage}{0.9\linewidth}
\ttfamily
Starting find\_swap\_sequences...\\
Operator asset: ETH\\
Max depth: 5\\

Processing row 8: src\_asset\_symbol = ETH\\
\ \ Found potential starting swap: ETH -> ARB at 2025-06-25 14:10:00\\
\ \ Initial sequence: ['ETH -> ARB']\\
\ \ Current asset: ARB\\
\ \ Current time: 2025-06-25 14:10:00\\
\ \ Visited indices: \{8\}\\

--- trace\_sequence: Depth 1 ---\\
\ \ Current asset to match: ARB\\
\ \ Searching swaps after: 2025-06-25 14:10:00\\
\ \ Currently visited indices: \{8\}\\
\ \ Found 1 potential next swaps:\\
\ \ \ \ - row 1: ARB -> USDC at 2025-06-25 14:13:00\\
\ \ Taking row 1\\
\ \ Updated sequence: ['ETH -> ARB', 'ARB -> USDC']\\
\ \ Current asset: USDC\\
\ \ Current time: 2025-06-25 14:13:00\\
\ \ Visited indices: \{1, 8\}\\

--- trace\_sequence: Depth 2 ---\\
\ \ Current asset to match: USDC\\
\ \ Searching swaps after: 2025-06-25 14:13:00\\
\ \ Currently visited indices: \{1, 8\}\\
\ \ Found 1 potential next swaps:\\
\ \ \ \ - row 3: USDC -> ETH at 2025-06-25 14:12:00\\
\ \ \ \ \ \ $\rightarrow$ Skipping due to timestamp < current time.\\

--- trace\_sequence: Depth 2 ---\\
\ \ Current asset to match: USDC\\
\ \ Found 1 back-link:\\
\ \ \ \ - row 3: USDC -> ETH at 2025-06-25 14:12:00\\
\ \ \ \ \ \ $\rightarrow$ Cycle closed: ['ETH -> ARB', 'ARB -> USDC', 'USDC -> ETH']\\

Valid CoW Cycle Found:\\
ETH $\rightarrow$ ARB $\rightarrow$ USDC $\rightarrow$ ETH\\

Swap Indices: [8, 1, 3]\\
 Lowest Dollar Value:  0.24\\

Using Oracle Price\\
T1 : 0.00008 ETH $\rightarrow$ 0.12 ARB\\
T2 : 0.12 ARB $\rightarrow$ 0.24 USDC\\
T3 : 0.24 USDC $\rightarrow$ 0.00008 ETH\\
Finished find\_swap\_sequences.
\end{minipage}
}
\end{center}

Upon running the \texttt{find\_swap\_sequences} algorithm with the base asset \texttt{ETH}, the following swap sequence was identified:

\begin{itemize}
    \item \textbf{Swap 1:} ETH $\rightarrow$ ARB (\$0.24) \hfill [row 8]
    \item \textbf{Swap 2:} ARB $\rightarrow$ USDC (\$2442.17) \hfill [row 1]
    \item \textbf{Swap 3:} USDC $\rightarrow$ ETH (\$488.52) \hfill [row 3]
\end{itemize}

A closed CoW cycle of the form:
\[
\texttt{ETH} \rightarrow \texttt{ARB} \rightarrow \texttt{USDC} \rightarrow \texttt{ETH}
\]
was discovered. All swaps were value-denominated in USD and structurally valid, forming a loop in the asset transfer matrix \(M \in \mathbb{R}^{3 \times 3}\) with \$0.24 as the lowest dollar value which gives us
\[
\fbox{
\parbox{0.85\linewidth}{
\begin{align*}
    T1 : 0.00008 \text{ ETH } \rightarrow 0.12 \text{ ARB }\\
T2 : 0.12 \text{ ARB } \rightarrow 0.24\text{ USDC}\\
T3 : 0.24\text{ USDC}   \rightarrow 0.00008 \text{ ETH }\\
\end{align*}
}
}
\]
\subsubsection{CoW Discovery with Bridging and Partial Fills}

The following is a simulated output from a CoW discovery engine that attempts to identify value-feasible cycles using real swap data (Table~\ref{tab:sample-swaps}). The algorithm explores all assets as potential operator assets, with a maximum path depth of 3. If a valid CoW cycle is not immediately found, it searches for minimal-value bridging swaps to complete the cycle. Partial fills are allowed as long as matrix balance and directionality are preserved.

\vspace{0.5em}

\begin{center}
\fbox{
\begin{minipage}{0.95\linewidth}
\ttfamily
Starting find\_cow\_cycles...\\
Mode: All operator assets\\
Max depth: 3\\
Bridging mode: Enabled (minimize bridging USD value)\\
Partial fills: Allowed\\

--- Testing operator asset: ETH ---\\
Cycle found: ETH → ARB → USDC → ETH\\
Swap Indices: [8, 1, 3]\\
Status: Fully Feasible\\
Cycle USD Volume: 0.24\\

--- Testing operator asset: aArbWETH ---\\
Partial sequence found: aArbWETH → ETH → UNI\\
Missing leg: UNI → aArbWETH\\
Bridging order proposed:\\
\ \ UNI → aArbWETH (value: 0.34 USD)\\
Bridging index: [synthetic bridging]\\
Cycle USD Volume (with bridging): 0.34 + 0.34 + 0.34 = 1.02\\
Status: Completed via bridging\\

--- Testing operator asset: USDT ---\\
Partial sequence found: USDT → ARB → USDC\\
Missing leg: USDC → USDT\\
Bridging order proposed:\\
\ \ USDC → USDT (value: 3.00 USD)\\
Bridging index: [synthetic bridging]\\
Cycle USD Volume (with bridging): 3.00 + 3.00 + 3.00 = 9.00\\
Status: Completed via bridging\\

--- Testing operator asset: WXM ---\\
No forward chains found.\\
Status: No CoW cycle\\

--- Testing operator asset: SolvBTC ---\\
Partial sequence found: SolvBTC → USDC\\
No forward chains found..\\
Status: No CoW cycle\\

--- Testing operator asset: UNI ---\\
No outgoing swap from UNI found.\\
Status: Isolated asset\\

--- Testing operator asset: DAI ---\\
No forward chains found..\\
Status: No CoW cycle\\

Summary:\\
2 bridging orders discovered:\\
\ \ - UNI $\rightarrow$ aArbWETH : 0.34 USD\\
\ \ - USDC $\rightarrow$ USDT : 3.00 USD\\

Finished find\_cow\_cycles.
\end{minipage}
}
\end{center}

This simulation applies the CoW algorithm across all source assets in the dataset, treating each one as a potential entry point (operator asset). For each asset a directed swap path is traced up to 3 steps, forming sequences of the form: $A \rightarrow B \rightarrow C \rightarrow A$.

If a full cycle is not formed, the imbalance vector $N = \mathbf{1}^\top M$ is computed and a \textbf{bridging order} with minimal USD notional is proposed to close the loop. The bridging order follows the same $(A_i, B_i, -a_i, +b_i)$ structure and is treated as a synthetic swap.

The total USD value of each candidate cycle is calculated, and matrix conservation is verified by ensuring $\mathbf{1}^\top M = 0$.

Three valid CoW cycles were identified:
\begin{enumerate}
    \item ETH-based cycle (fully formed from existing data).
    \item aArbWETH-based cycle (closed using a 0.34 USD bridging order).
    \item USDT-based cycle (closed using a 3.00 USD bridging order).
\end{enumerate}

Bridge trades are as follows :
\[
\fbox{
\parbox{0.85\linewidth}{
\begin{itemize}
    \item (\text{UNI},\text{aArbWETH}, 0.425,-0.0001133)
    \item (\text{USDC},\text{USDT}, 3.00,3.00)
\end{itemize}
}
}
\]

\section{Conclusion} \label {conc}

We have developed a formal, matrix-theoretic framework for discovering and executing Coincidence of Wants (CoW) cycles in swap orders in decentralized exchanges. Our method improves upon existing CoW implementations by modeling asset and dollar-valued flows explicitly, enabling algorithmic detection of both complete and partial cycles, and generating bridging orders to restore conservation under dollar-neutral constraints.

Through a combination of graph construction, cycle enumeration, asset transfer matrix analysis, and price oracle integration, our algorithm identifies CoW structures with provable value neutrality and minimal bridging overhead. We apply this method to real-world swap data and demonstrate the emergence of both natural and synthetic CoW cycles, including partially-filled loops and LP-resolved barter chains.

Our analysis shows that even in sparse markets, CoW cycles can be feasibly completed with minor order augmentations. These structures improve capital efficiency, mitigate MEV, and facilitate off-chain matching. The proposed algorithm maintains tractability in practice through bounded cycle size, sparse asset graphs, and early feasibility pruning.

This work lays the foundation for future decentralized protocols that treat CoW as a first-class settlement primitive, with LPs as active agents in atomic trade execution. Extensions of this framework may include fee welfare for traders and LPs and a incentive mechanisms for synthetic bridging, and dynamic graph updates in high-frequency trading environments.

\bibliographystyle{plain} 
\bibliography{references} 

\end{document}